\documentclass[preprintnumbers,prd,floatfix,nofootinbib,superscriptaddress]{revtex4}

\usepackage[paperwidth=215.9mm,paperheight=279.4mm,centering,hmargin=2cm,vmargin=2.5cm]{geometry}

\usepackage[tbtags]{amsmath}  
\usepackage{tabularx}
\usepackage{amssymb}          
\usepackage{bm}               
\usepackage{graphicx}         
\usepackage[export]{adjustbox}
\usepackage{hhline,multirow}  
\usepackage{dcolumn}          
\usepackage{slashed}
\usepackage{datetime}
\usepackage[dvipsnames]{xcolor}
\usepackage{slashed}

\usepackage{amscd}            
\usepackage{epsfig}
\usepackage{dcolumn}          
\usepackage[dvipsnames]{xcolor}
\usepackage[normalem]{ulem}
\usepackage{mathtools}
\usepackage{comment}
\usepackage[utf8]{inputenc}

\RequirePackage{color}
\RequirePackage[colorlinks=true
,urlcolor=black
,anchorcolor=black
,citecolor=black
,filecolor=black
,linkcolor=black
,menucolor=black
,pagecolor=black
,linktocpage=black
,pdfproducer=medialab
,pdfa=true
]{hyperref}

\newcommand{\nocontentsline}[3]{}
\newcommand{\tocless}[2]{\bgroup\let\addcontentsline=\nocontentsline#1{#2}\egroup}

\newcommand{\al}{\alpha}

\newcommand{\dd}{\text{d}}

\newcommand{\st}{{\scriptscriptstyle T}}

\graphicspath{{/Users/asignori/Downloads/TMD_special_issue/Figures/pp_quarks/cute/evaluations/}
{/Users/asignori/Downloads/TMD_special_issue/Figures/pp_quarks/w_mass_extraction/pythia8_yoda/plots/}
{/Users/asignori/Downloads/TMD_special_issue/Figures/pp_quarks/dyqt}
{/Users/asignori/Downloads/TMD_special_issue/Figures/pp_quarks/plots}}

\allowdisplaybreaks[2]
\DeclareUnicodeCharacter{2212}{-}

\begin{document}

\title{Non-perturbative uncertainties on the transverse momentum distribution of electroweak bosons and on the determination of the $W$ boson mass at the LHC}

\newcommand*{\PaviaU}{Dipartimento di Fisica, Universit\`a di Pavia,
  via Bassi 6, I-27100 Pavia, Italy}\affiliation{\PaviaU}
\newcommand*{\InfnPavia}{INFN, Sezione di Pavia,
  via Bassi 6, I-27100 Pavia, Italy}\affiliation{\InfnPavia}
\newcommand*{\ANL}{Physics Division, Argonne National Laboratory \\ 9700 S. Cass Avenue, Lemont, IL 60439 USA}\affiliation{\ANL}

\author{Giuseppe Bozzi}
\thanks{giuseppe.bozzi@unipv.it - \href{https://orcid.org/0000-0002-2908-6077}{ORCID: 0000-0002-2908-6077}}
\affiliation{\PaviaU}\affiliation{\InfnPavia}

\author{Andrea Signori}
\thanks{asignori@anl.gov - \href{https://orcid.org/0000-0001-6640-9659}{ORCID: 0000-0001-6640-9659}} 
\affiliation{\ANL}

\begin{abstract}
In this contribution we present an overview of recent results concerning the impact of a possible flavour dependence of the intrinsic quark transverse momentum on electroweak observables. In particular, we focus on the $q_\st$ spectrum of electroweak gauge bosons produced in proton-proton collisions at the LHC and on the direct determination of the $W$ boson mass. We show that these effects are comparable in size to other non-perturbative effects commonly included in phenomenological analyses, and should thus be included in precise theoretical predictions for present and future hadron colliders.
\end{abstract}

\date{\today}
\maketitle
\tableofcontents
\newpage

\section{Introduction}
\label{s:intro_wmass}

Electroweak precision observables are interesting benchmarks to test the limits of the Standard Model and to discriminate between different scenarios for new physics. The mass of the $W$ boson, $m_W$, is an example of such an observable. 

The Standard Model prediction for the $W$ boson mass from the global fit of the electroweak parameters ($m_W=80.356 \pm 0.008$ GeV)~\cite{Baak:2014ora} has a very small uncertainty that represents a natural target for the precision of the experimental measurements of $m_{W}$ at hadron colliders.

\begin{figure}[hbt!]
\begin{center}
\includegraphics[width=0.5\textwidth]{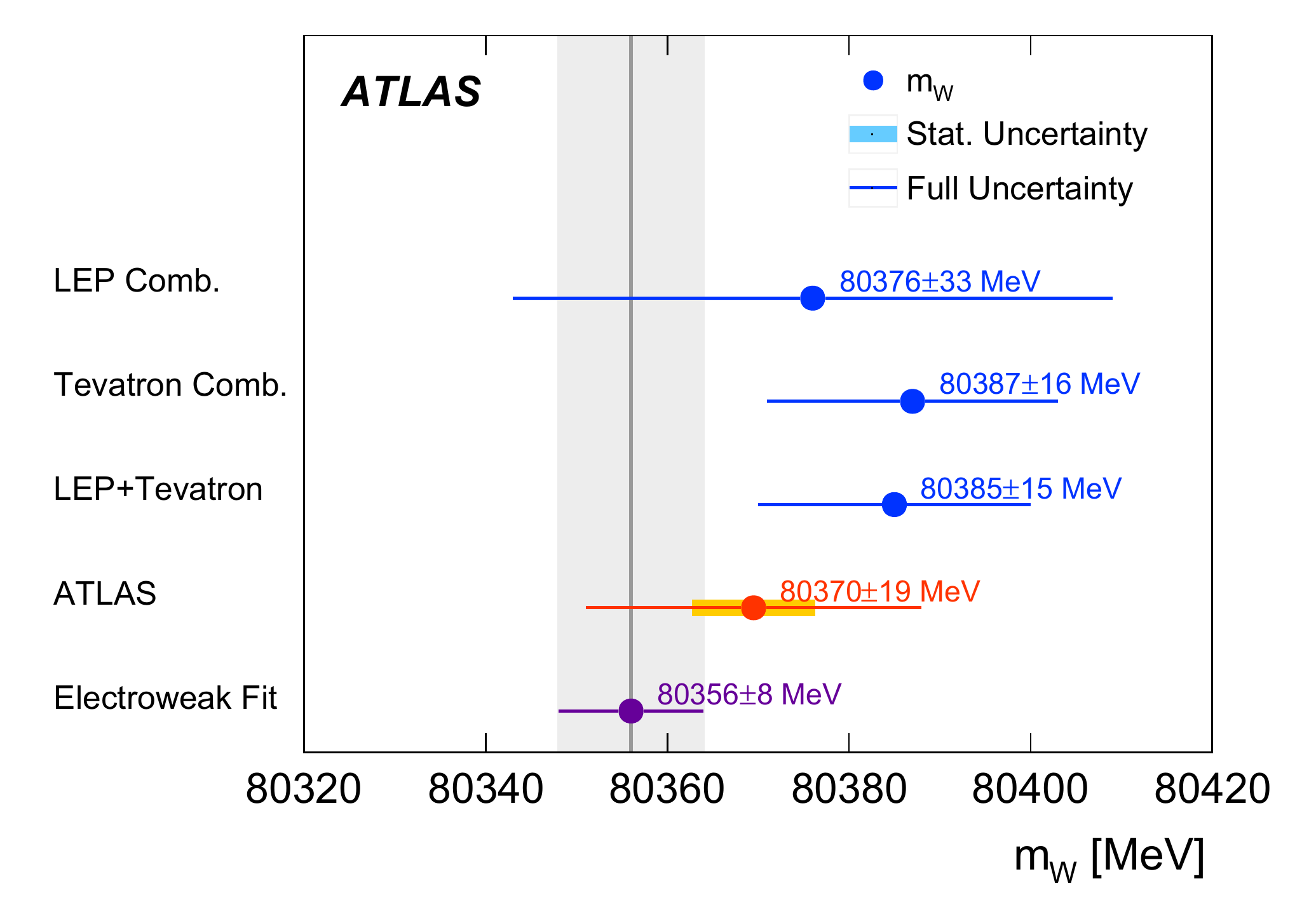} 
\end{center}
\caption{Overview of the measurements of the $W$ boson mass. The indirect determination via the electroweak fit sets the precision for the measurements via direct determinations. Figure from Ref.~\cite{Aaboud:2017svj}.}
\label{f:wmass_det}
\end{figure}

Direct measurements of $m_W$ at hadronic colliders have been performed at the Tevatron $p\bar{p}$ collider with the {\tt D0}~\cite{D0:2013jba} and {\tt CDF}~\cite{Aaltonen:2013vwa} experiments, and at the LHC $pp$ collider with the {\tt ATLAS}~\cite{Aaboud:2017svj} experiment, with a total uncertainty of 23 MeV, 19 MeV and 19 MeV, respectively. The current world average, based on these measurements and the ones performed at LEP, is $m_W=80.379 \pm 0.012$ GeV~\cite{Tanabashi:2018oca}. Fig.~\ref{f:wmass_det} presents an overview of these measurements compared to the electroweak global fit. The CPT theorem~\cite{Weinberg:1995mt,Peskin:1995ev} implies that the mass and lifetime of a particle and its anti-particle are the same. The {\tt ATLAS} measurement of the $W^{+}$ and $W^{-}$ mass difference yields: $m_{W^+} - m_{W^-} = -29 \pm 28$ MeV~\cite{Aaboud:2017svj}. 
The experimental determinations are based on a template-fit procedure applied to differential distributions of the $W$ decay products: in particular, the transverse momentum of the final lepton, $p_\st^\ell$, the transverse momentum of the neutrino $p_\st^\nu$ (only at the Tevatron), and the transverse mass $m_\st$ of the lepton pair (where $m_\st=\sqrt{2\;p_\st^\ell\;p_\st^\nu\;(1-\cos(\phi_{\ell}-\phi_{\nu}))}$, with $\phi_{\ell , \nu}$ being the azimuthal angles of the lepton and the neutrino, respectively). The transverse momentum of the lepton pair, though not directly used in the template-fit procedure, is relevant for reweighing purposes (see, for instance, Sec. 6 of Ref.~\cite{Aaboud:2017svj}).

At leading order the $W$ boson is produced with zero transverse momentum ($q_{\st}^W$), but perturbative and non-perturbative corrections give rise to non-vanishing values of $q_{\st}^W$. While perturbative and flavour-independent non-perturbative corrections have received much attention and reached a high level of accuracy (see, for instance, Ref.~\cite{Alioli:2016fum,CarloniCalame:2016ouw} and references therein), a possible flavour dependence of the intrinsic transverse momentum ($k_{\st}$) of the initial state partons has been less investigated.

\begin{figure}[hbt!]
\begin{center}
\includegraphics[width=0.47\textwidth]{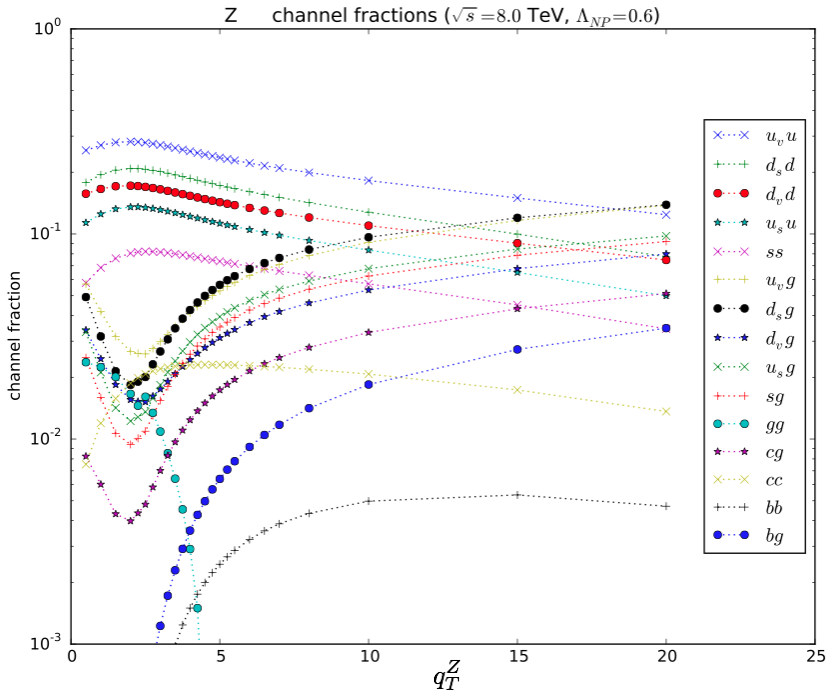} 
\\
\includegraphics[width=0.47\textwidth]{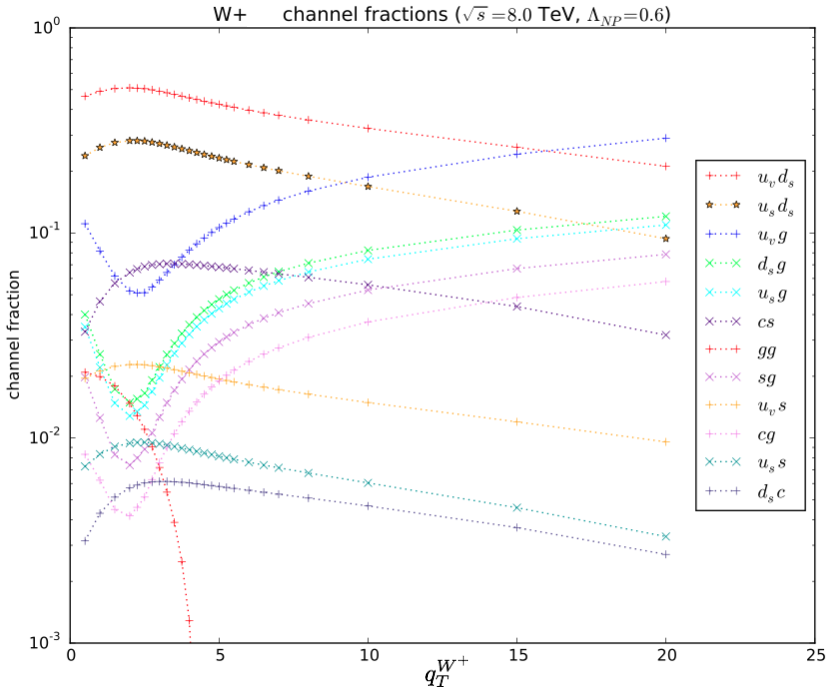}
\\
\includegraphics[width=0.47\textwidth]{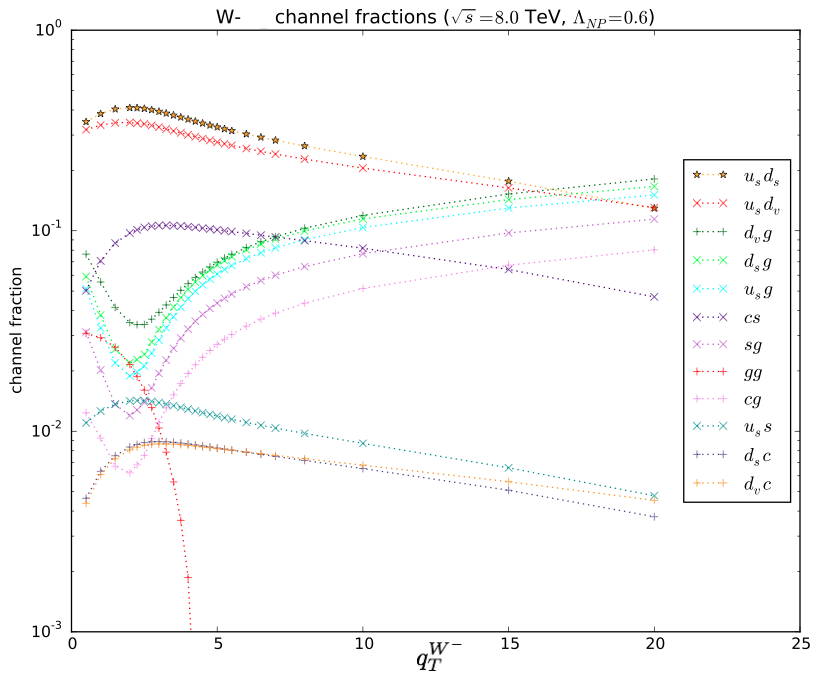}
\end{center}
\caption{
From top to bottom: the decomposition in flavour channels of the cross section for $Z$, $W^+$, $W^-$ production differential with respect to the transverse momentum of the produced electroweak boson $q_\st^V$, $V=Z, W^+, W^-$. The cross section is calculated by means of {\tt CuTe}~\cite{Becher:2011xn} at LHC $\sqrt{s} = 8$ TeV. The non-perturbative correction is implemented as a flavour-independent Gaussian smearing, governed by the parameter $\Lambda_{NP}$ (see Ref.~\cite{Becher:2011xn} and App.~\ref{app:con_NP_par}). The channels add to one.  
}
\label{f:flav_channels}
\end{figure}

In Fig.~\ref{f:flav_channels} we examine the decomposition in flavour channels of the cross section for $Z$ and $W^\pm$ production differential with respect to $q_\st^V$, $V=Z, W^\pm$. A non-trivial interplay among the different flavours and the gluon is observed. The role of the gluon becomes increasingly important at larger values of the transverse momentum. In the region of the peak, instead, the dominant channels involve combinations of $u_{val}$, $d_{val}$, $\bar{u}$ and $\bar{d}$ (where $a = a_{val} + a_{sea}$ and $\bar{a} = a_{sea}$). For this reason, we consider important to study the impact of flavour-dependent effects on the production of electroweak bosons and on the determination of $m_W$. 

In this contribution we give an overview of selected studies related to flavour-dependent effects, focusing in particular on the results obtained in Ref.~\cite{Signori:2016lvd,Bacchetta:2018lna}, showing that they can be non-negligible compared to other sources of theoretical uncertainty and should thus be included in precision physics programs at hadron colliders.

\section{Formalism}
\label{s:formalism_wmass}

In processes with a hard scale $Q$ and a measured transverse momentum $q_\st$, for instance the mass and the transverse momentum of an electroweak boson produced in hadronic collisions, we can distinguish three regions: a small $q_{\st}$ region ($q_\st \ll Q$), where large logarithms of $q_{\st}/Q$ have to be properly resummed; a large $q_{\st}$ region ($q_{\st} \gtrsim Q$), where fixed-order perturbation theory provides reliable results; and an intermediate region, where a proper matching procedure between all-order resummed and fixed-order contributions is necessary. For a concise discussion (and for additional relevant references) on the development of the different frameworks available to resum the logs of $q_{\st}/Q$ and on their matching to fixed-order perturbative calculations, we refer the reader to Ref.~\cite{Catani:2013tia,Boglione:2014oea,Collins:2016hqq,Collins:2017oxh,Echevarria:2018qyi}. 

In the Transverse-Momentum-Dependent (TMD) factorisation framework~\cite{Collins:2011zzd}, the unpolarized TMD Parton Distribution Function (TMD PDF) for a parton with flavour $a$, carrying a fraction $x$ of longitudinal momentum at a certain scale $Q^2$, can be written in $b_\st$-space (where $b_\st$ is the variable Fourier-conjugated to the partonic transverse momentum $k_\st$) as :
\begin{equation}
\widetilde{f}_1^a (x,  b_\st; Q^2) = \sum_{i=q,\bar q,g} \bigl( C_{a/i} \otimes f_1^i \bigr) (x,b_\st,\mu_b^2) 
\  e^{S (\mu_b^2, Q^2)} e^{g_{K}(b_\st,\bm{\lambda})\ln(Q^{2}/Q_{0}^{2})} \widetilde{f}_{{\rm NP}}^a (b_\st,\bm{\lambda^\prime}) \ ,
\label{e:TMDevol1}
\end{equation}
where $\mu_{b}$ is the $b_\st$-dependent scale at which the collinear parton distribution functions are computed and $Q_{0}$ is a hadronic mass scale. Eq.~\eqref{e:TMDevol1} is a generic schematic implementation of the perturbative and non-perturbative components of a renormalized TMD PDF. Depending on the  chosen perturbative accuracy, $S$ includes the UV-anomalous dimension of the TMD PDF and the Collins-Soper kernel. Also, in principle the TMD PDF depends on two kinds of renormalization scales, related to the renormalization of UV and light-cone divergences. Here we specify their initial and final values as $\mu_b$ and $Q$ respectively. Moreover, the perturbative scales can be chosen in position or momentum space~\cite{Bozzi:2010xn,Becher:2011xn,Catani:2015vma,Bacchetta:2015ora,Bacchetta:2017gcc,Kang:2017cjk}. For the implementation of all these details, we refer the reader to the description of the public codes that we are going to discuss.

The $C$ coefficients in Eq.~\eqref{e:TMDevol1}, also called Wilson coefficients for the TMD distribution, are calculable in perturbation theory and are presently known at order $\alpha_{s}^{2}$ in the unpolarized case~\cite{Echevarria:2016scs,Collins:2017oxh}. They are convoluted with the corresponding collinear parton distribution functions $f_1^{i}$ according to 
\begin{equation}
\bigl( C_{a/i} \otimes f_1^i \bigr) (x, b_\st, \mu_b^2) =
 \int_x^1 \frac{du}{u}\  
       C_{a/i} \Big( \frac{x}{u}, b_\st, \alpha_s\big(\mu_b^2\big)  \Big) \  
       f_1^i (u; \mu_b^2) \  , 
\label{e:WC1}
\end{equation}

The perturbative part of the evolution, the $S$ factor in Eq.~\eqref{e:TMDevol1}, can be written as: 
\begin{equation}
\label{e:sud}
S(\mu_b^2,Q^2) = \int_{\mu_b^2}^{Q^2}{d\mu^2\over \mu^2} \gamma_F[\alpha_s(\mu^2),Q^2/\mu^2] - K(b_\st;\mu_b^2) \log \frac{Q^2}{\mu_b^2} \ .
\end{equation}
It involves, in principle, the UV-anomalous dimension $\gamma_F$ and the Collins-Soper kernel $K$, which can be decomposed as:
\begin{equation} 
\gamma_F[\alpha_s(\mu^2),Q^2/\mu^2]=-
\bigg[ \sum_{k=1}^{\infty}A_k \bigg(\frac{\alpha_s(\mu^2)}{4\pi} \bigg)^k \bigg] \ln\bigg({Q^2\over \mu^2}\bigg) + 
\sum_{k=1}^{\infty}B_k \bigg(\frac{\alpha_s(\mu^2)}{4\pi} \bigg)^k 
\ , 
\ \ \ \  K(b_\st,\mu_b^2) = \sum_{k=1}^{\infty} d_k \bigg(\frac{\alpha_s(\mu^2)}{4\pi} \bigg)^k \ .
\label{e:Sudakov} 
\end{equation} 
The $A_{k}$ and $B_{k}$ coefficients are known up to NNNLL (at least, their numerical value) and the integration of the Sudakov exponent in Eq.~\eqref{e:Sudakov} can be done analytically up to NNNLL (for the complete expressions see, e.g., Ref.~\cite{Bozzi:2005wk,Echevarria:2012pw,Bizon:2018foh}). The perturbative coefficients of the kernel $K$ are also known analytically up to NNNLL.  

A well known problem in the implementation of the QCD evolution of transverse-momentum-dependent distributions (TMDs) is the divergent behaviour at large $b_\st$ caused by the QCD Landau pole. Two common prescriptions to deal with this divergence consist in replacing $b_\st$ with a variable that saturates at a certain $b_{\st\rm max}$, as suggested by the CSS formalism~\cite{Collins:2011zzd,Aybat:2011zv}, or perform the $b_\st$ integration on the complex plane in such a way that the Landau pole is never reached~\cite{Laenen:2000de}. On the other hand, also the small $b_\st$ region needs to be regularized, in order to eliminate unjustified contributions from the evolution of TMDs in the intermediate and large $q_{\st}$ regions and to recover the expression for the cross section in collinear factorisation upon integration over $q_\st$. Several prescriptions exist~\cite{Bozzi:2005wk,Boer:2014tka,Boer:2015uqa,Collins:2016hqq,Bacchetta:2017gcc} also in this case.

Two intrinsically non-perturbative factors are introduced in Eq.~\eqref{e:TMDevol1} in order to account for the large $b_\st$ behavior. The first one is named $g_K(b_\st;\bm{\lambda})$ in the TMD/CSS literature~\cite{Collins:2011zzd}. It embodies the flavour-independent non-perturbative part of the evolution. The second one, $\widetilde{f}_{{\rm NP}}^a (b_\st; \bm{\lambda}^\prime)$, accounts for a kinematic- and flavour-(in)dependent intrinsic transverse momentum of the parton with flavour $a$. The $\bm{\lambda}$ and $\bm{\lambda^\prime}$ are (vectors of) nonperturbative parameters that can be fit to data. The $\bm{\lambda^\prime}$ parameters are related to the quantity $\langle \bm{k}_\st^2 \rangle_a$. For example, in case of a simple Gaussian functional form, $e^{-\lambda^\prime b_\st^2}$, we have $\lambda^\prime = \langle \bm{k}_\st^2 \rangle_a / 4$. For both the nonperturbative factors $g_K$ and $\widetilde{f}_{{\rm NP}}^a$, several implementations have been discussed, see, e.g., Ref.~\cite{Scimemi:2017etj,Bacchetta:2017gcc} and references therein. In particular, a kinematic- and flavour-dependent Gaussian parametrisation has been proposed in Ref.~\cite{Signori:2013mda,Bacchetta:2015ora}. 

The studies that we discuss make use of three different computational tools: \href{http://cute.hepforge.org/}{{\tt CuTe}}~\cite{Becher:2011xn}, \href{http://pcteserver.mi.infn.it/~ferrera/dyqt.html}{{\tt DyqT}}~\cite{Bozzi:2010xn} and \href{http://theory.fi.infn.it/grazzini/dy.html}{{\tt DYRes}}~\cite{Catani:2015vma}.
{\tt CuTe} implements the SCET formalism, where the resummation is performed in terms of factorisation formulae that involve Soft Collinear Effective Theory operators and matching coefficients. It gives the transverse momentum spectrum of on-shell electroweak bosons at NLO (${\cal O}(\alpha_{s}$)) accuracy in the $C$ Wilson coefficients and at NNLL in the Sudakov exponent\footnote{{\tt CuTe} is labelled NNLL in the SCET language but NNLL$^\prime$ in standard pQCD language. The accuracy of NNLL$^\prime$ is considered lower than that of the full NNLL, in which Wilson coefficients are computed at NNLO.}.

{\tt DyqT} and {\tt DYRes} are based on Ref.~\cite{Bozzi:2010xn,Catani:2015vma} and perform soft gluon resummation in $b_\st$-space. The first computes the $q_{\st}$ spectrum of an electroweak boson produced in hadronic collisions. The second also provides the full kinematics of the vector boson and of its decay products, allowing for the application of arbitrary cuts on the final-state kinematical variables and giving differential distributions in form of bin histograms. The accuracy of both codes is up to NNLL in the resummed part, and up to NLO (${\cal O}(\alpha_{s}^{2})$) at large $q_\st$.

A simple Gaussian parametrisation of the non-perturbative effects is present in these codes, as in most of the computational tools used to analyse the electroweak observables relevant for the determination of the $W$ boson mass. A single non-perturbative parameter, $g_{NP}$, usually encodes both the (flavour-independent) effect of $g_K$ and the distribution in the (potentially flavour-dependent) intrinsic transverse momentum\footnote{{\tt ResBos}~\cite{Landry:2002ix} is a counter-example, but it does not account for the flavour dependence of the intrinsic transverse momentum.}: 
\begin{equation}
 e^{- g_{NP}  b_T^2}\ \equiv \ e^{2g_K(b_T;\bm{\lambda}) \ln (Q^2/Q_0^2)}\ 
\widetilde{f}_{{\rm NP}}^a (b_\st; \bm{\lambda}^\prime) \ 
\widetilde{f}_{{\rm NP}}^{a^\prime} (b_\st; \bm{\lambda}^\prime) \ .
\label{e:gauss_NP_TMDPDF}
\end{equation}

The values of the non-perturbative parameters used in fitting the $W$ boson mass are usually obtained through fits on $Z$ production data~\cite{Landry:2002ix}, for which the relevant partonic channels are of the type $q_i\bar{q}_i$, and then used to predict $W^\pm$ production, despite the process being sensitive to different partonic channels, $q_i\bar{q_j}$ ($i \neq j$). This procedure essentially neglects any possible flavour dependence of the intrinsic partonic transverse momentum. 

In order to introduce the flavour dependence, one can simply decompose $g_{NP}$ in the LHS of Eq.~\eqref{e:gauss_NP_TMDPDF} into the sum $g_{NP}^a + g_{NP}^{a'}$, where the flavour indices span the range $a,a^\prime =u_v, u_s, d_v, d_s, s, c, b, g$ (the subscripts referring to the valence and sea components, respectively), additionally disentangling the non-perturbative contribution to the evolution and the intrinsic transverse momentum distribution. 
Thus, for each parton with flavour $a$, the nonperturbative contribution $\widetilde{f}_{{\rm NP}}^a$ and $g_K$ in Eq.~\eqref{e:TMDevol1},~\eqref{e:gauss_NP_TMDPDF} are included in the corresponding term in the flavour sum of the TMD factorisation formula. 
More details regarding the non-perturbative parameters in the codes under consideration have been collected in App.~\ref{app:con_NP_par}.
 
\section{Effects on the $q_{\st}$ spectrum of the $W$}
\label{s:effects_peak_wmass}

The impact of a flavour-dependent intrinsic $\langle \bm{k}_{\st}^2 \rangle$ on the $q_\st$ spectrum of the electroweak bosons has been first studied in Ref.~\cite{Signori:2016lvd} and here we partly summarize the findings therein. Part of the analysis is devoted to the shifts induced in the position of the peak for the distribution in $q_\st^V$, $V = W^{+}, W^{-}$ and $Z$. Flavour-independent (f.i.) and flavour-dependent (f.d.) variations of the average intrinsic transverse momentum squared are considered, together with the uncertainties associated to other non-perturbative factors, such as the collinear PDFs, the renormalisation scale, and the value of the strong coupling constant. As justified in Sec.~\ref{s:intro_wmass} and Fig.~\ref{f:flav_channels}, it is assumed that the intrinsic transverse-momentum depends on five flavours only: $u_{v}, d_{v}, u_{s}, d_{s}, s$, where $s$ collectively refers to the strange, charm and bottom quarks and to the gluon.

The numerical results are obtained by means of a modified (i.e. flavour-dependent) version of {\tt CuTe}~\cite{Becher:2011xn}. Namely, the non-perturbative parameter $2\Lambda_{NP}^2$ (see App.~\ref{app:con_NP_par}), which corrects the whole cross section at large $b_\st$, is split into a sum of two flavour-dependent non-perturbative contributions, $\Lambda_{i,j}$, such that $\Lambda_{i}^2 + \Lambda_{j}^2 = 2\Lambda_{NP}^2$. This decomposition reabsorbs the non-perturbative contribution to QCD radiation into $\Lambda_{i,j}$. The flavour dependence of $\Lambda_{i,j}$ is compatible with the ratios fitted in Ref.~\cite{Signori:2013mda}. The goal is to combine flavour dependent parameters in such a way to respect the values of $\Lambda_{NP}$ fitted on the $Z$ data, generating at the same time different values $\Lambda_{i,j}$ to be used in the calculation of the differential cross section for $W^\pm$ (we refer the reader to Ref.~\cite{Signori:2016lvd} for the precise values of $\Lambda_{i,j}$ used in the study). 

The shifts (quantified in GeV) induced by different perturbative and non-perturbative contributions are summarized in Tab.~\ref{t:summ_shifts}. The renormalisation scale is varied between $1/2 \mu_c$ and $2 \mu_c$, with $\mu_c = q_\st + q^\star$, where $q^\star$ is a cutoff introduced in the SCET formalism to avoid the Landau pole~\cite{Becher:2011xn}. The scale in the hard part has not been varied. Regarding the impact of the collinear PDFs, the result shown in the table is the smallest interval which contains $68\%$ or $90\%$ of peak positions, computed for every member of the NNPDF3.0 set~\cite{Ball:2014uwa}. The strong coupling is varied by $\pm 0.003$ from the central value of 0.118. 
\begin{table}[h!]
\begin{center}
\hrule height 0.5pt\smallskip
\caption{Summary of the shifts in GeV induced on the peak position in $q_\st$ spectra of $W^\pm/Z$, generated by different effects. ``f.i.'' stands for flavour-independent, whereas ``f.d.'' for flavour-dependent. ``Max $W^\pm$'' effect indicates the maximum shift induced on the peak position of the $W^\pm$ $q_\st$ spectrum by flavour-dependent variations of $\langle \bm{k}_\st^2 \rangle$ that keep the peak of the $Z$ $q_\st$ spectrum unchanged. For the values of the flavour-dependent non-perturbative parameters we refer the reader to Ref.~\cite{Signori:2016lvd}.}
\label{t:summ_shifts}
\begin{tabular}{l|l l|l l|l l}
							&	$W^+$	&			&	$W^-$	&			&	$Z$		&\\
\cline{1-7}							
$\mu_R = \mu_c/2, 2 \mu_c$	&	$+0.30$	&	$-0.09$	&	$+0.29$	&	$-0.06$	&	$+0.23$	&	$-0.05$	\\
pdf ($68\%$ cl)				&	$+0.03$	&	$+0.03$	&	$+0.04$	&	$+0.00$	&	$+0.03$	&	$-0.02$	\\
pdf ($90\%$ cl)				&	$+0.03$	&	$-0.05$	&	$+0.06$	&	$-0.02$	&	$+0.05$	&	$-0.02$	\\
$\al_s = 0.118 \pm 0.003$		&	$+0.14$	&	$-0.12$	&	$+0.14$	&	$-0.14$	&	$+0.15$	&	$-0.15$	\\
f.i. $\langle \bm{k}_\st^2 \rangle = 1.0, 1.96$ 	&	$+0.16$ & $-0.16$	&	$+0.16$ & $-0.14$	&	$+0.16$ & $-0.15$	\\
f.d. $\langle \bm{k}_\st^2 \rangle$ (max $W^+$ effect)	&	$+0.09$ & &	& $-0.06$		&	$\pm 0$		\\
f.d. $\langle \bm{k}_\st^2 \rangle$ (max $W^-$ effect)	&	& $-0.03$	&	$+0.05$ &		&	$\pm 0$		
\end{tabular}
\end{center}
\hrule height 0.5pt\smallskip
\end{table}

The shift induced in the peak position from flavour-dependent $\langle \bm{k}_\st^2 \rangle$ is smaller than that induced by scale variation, $\al_s$ variation and flavour-independent $\langle \bm{k}_\st^2 \rangle$, but comparable in magnitude. It is also bigger than the uncertainty from the PDF set, which is the only other uncertainty where the shifts are not almost perfectly correlated between the three vector bosons. With flavour-dependent variations of $\langle \bm{k}_\st^2 \rangle$, the peaks of the $W^+$ and $W^-$ distributions shift in different directions. Since the $\langle \bm{k}_\st^2 \rangle$ parameters are selected under the constraint that the $Z$ $q_\st$-distribution is left unchanged (see Tab.~\ref{t:summ_shifts}), the channels for $W^+$ and $W^-$ move in different directions. 
The anticorrelation of the shifts between $W^+$ and $W^-$ is a peculiarity of the uncertainty generated by flavour-dependent variations of the intrinsic $k_{\st}$. This means that the uncertainty stemming from the non-perturbative hadron structure in the transverse plane can affect the determination of $m_{W^+}$ and $m_{W^-}$ in different ways. 
Indeed, this feature nicely emerges in the analysis summarized in Sec.~\ref{s:impact_wmass}. 

The analysis in Ref.~\cite{Signori:2016lvd} thus shows that the uncertainty on the peak position for $W^\pm$ bosons arising from the flavour dependence of the intrinsic transverse momentum is not negligible with respect to the other sources of theoretical uncertainties and comparable in magnitude with the uncertainties due to the collinear PDFs. \\

\begin{figure}[hbt!]
\begin{center}
\includegraphics[width=0.5\textwidth]{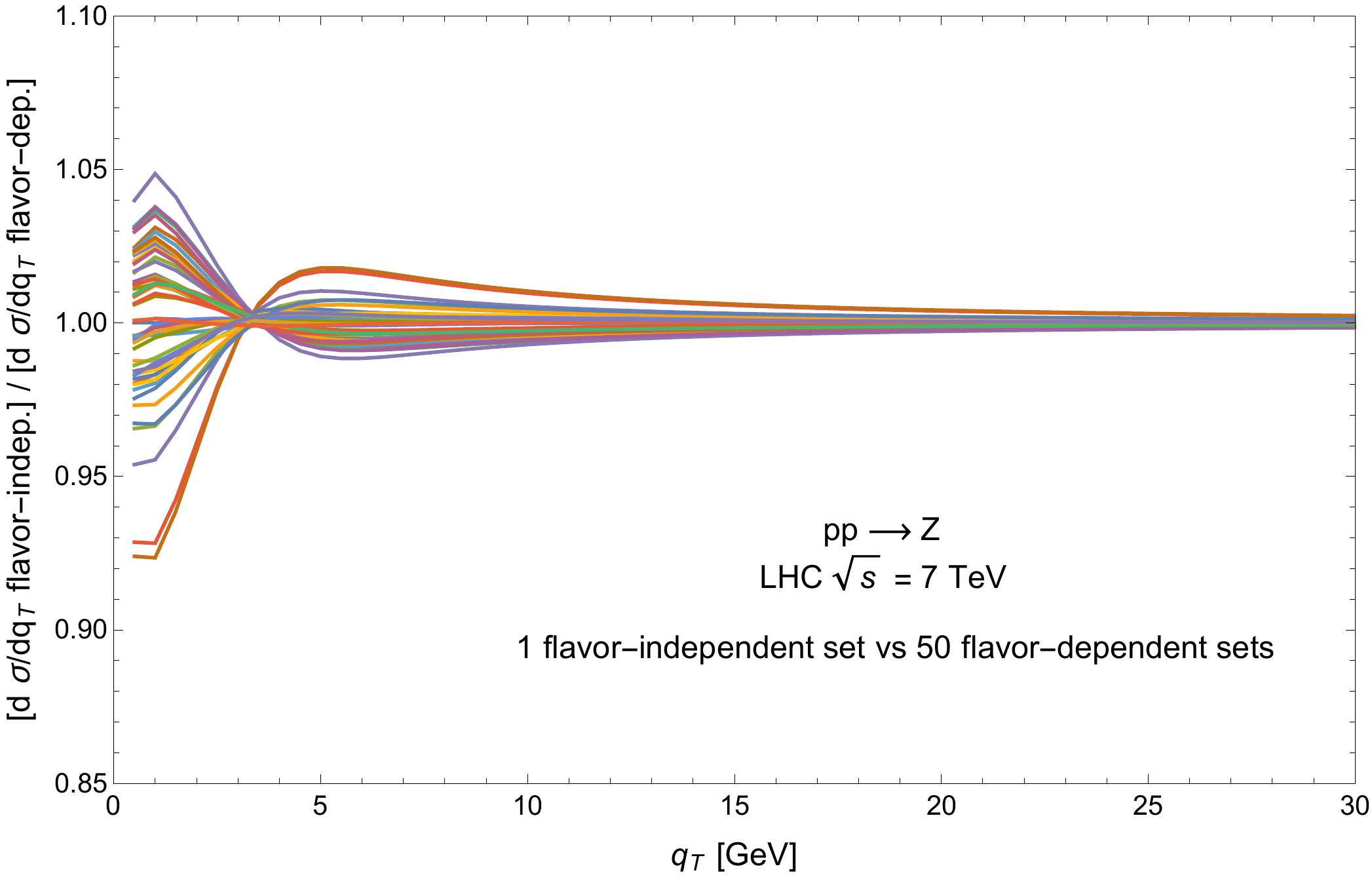} 
\\
\includegraphics[width=0.5\textwidth]{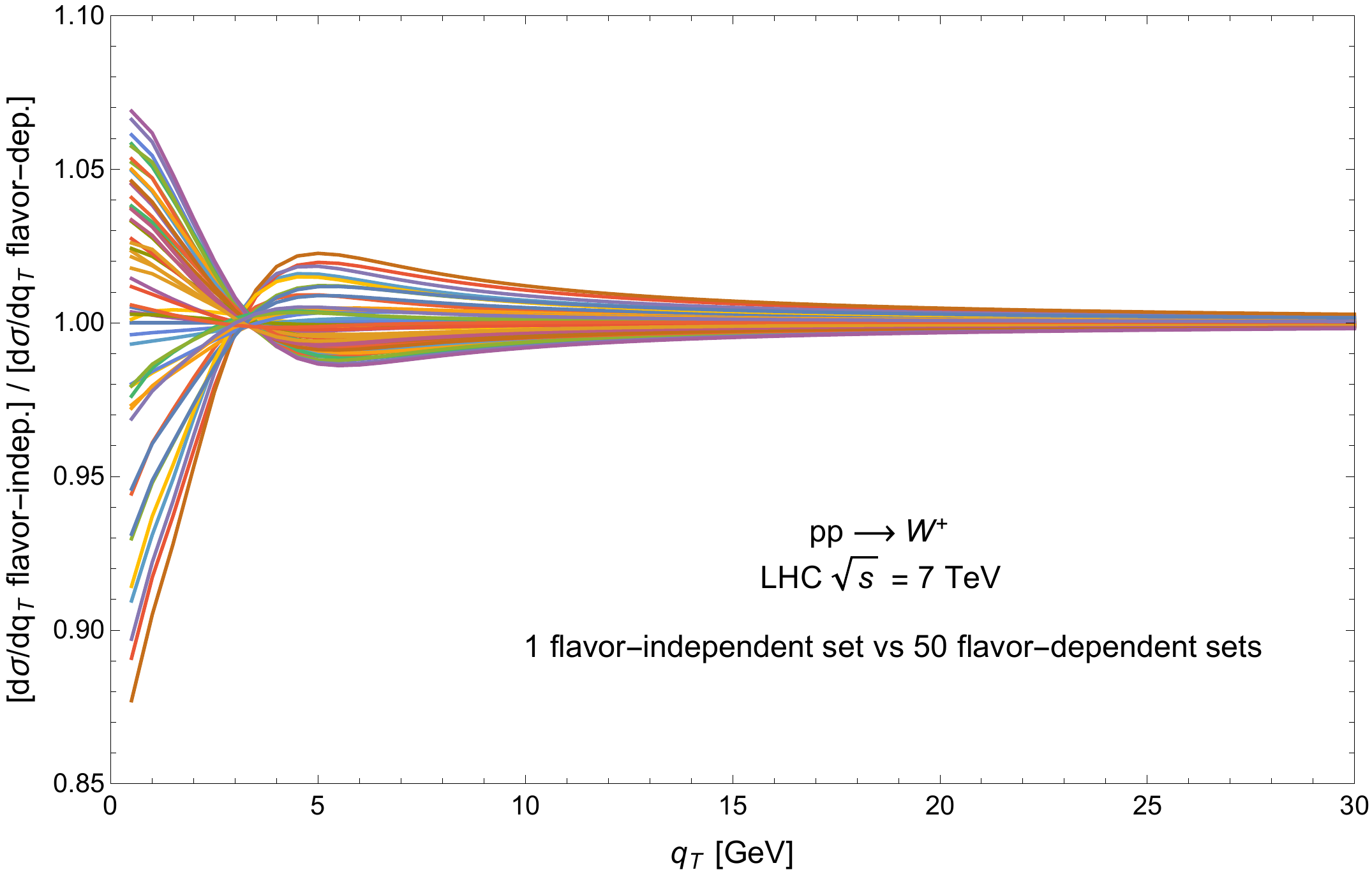}
\\
\includegraphics[width=0.5\textwidth]{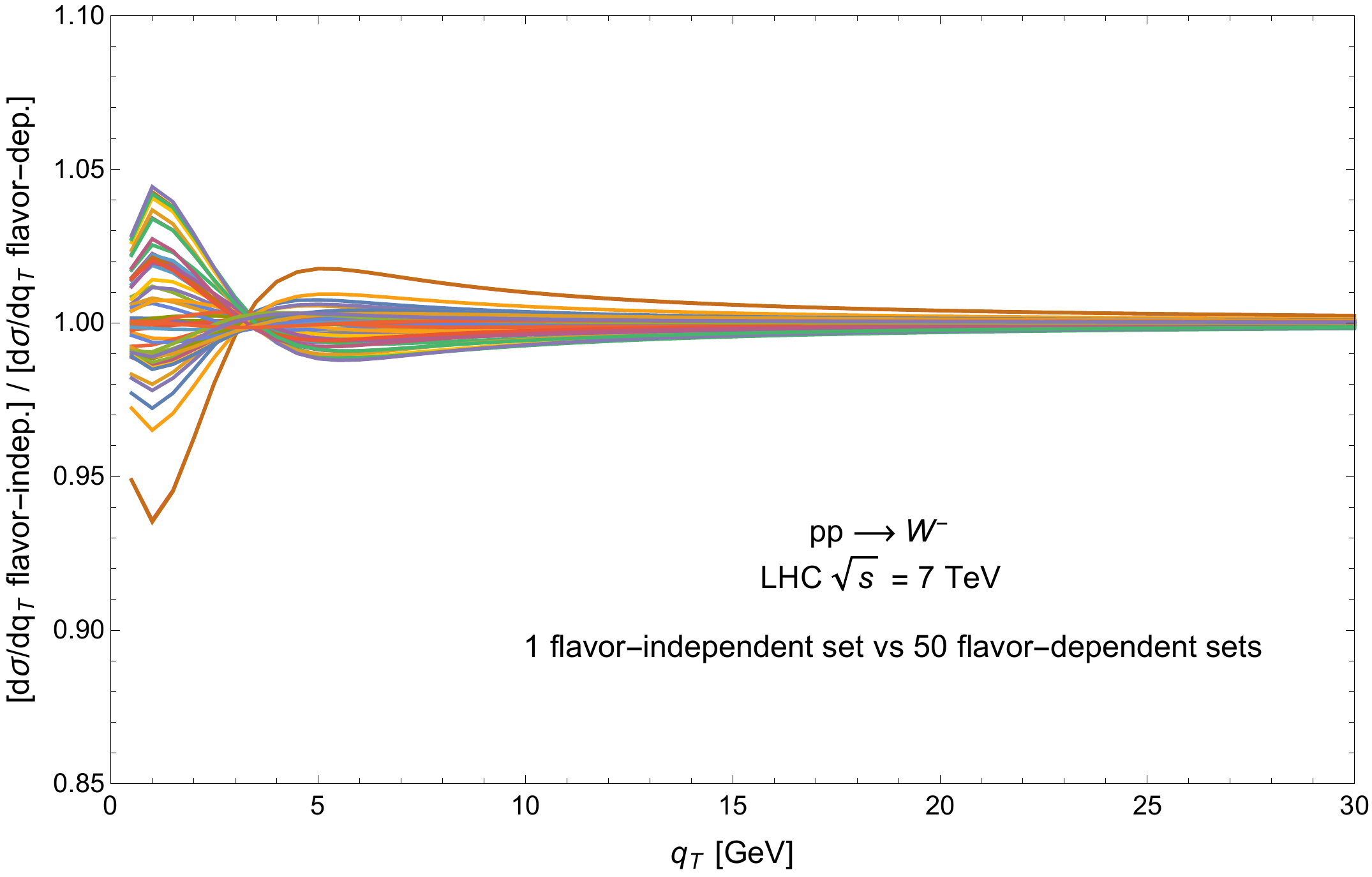}
\end{center}
\caption{
In these figures the ratio $\frac{d\sigma^V}{dq_\st}(f.i.) / \frac{d\sigma^V}{dq_\st}(f.d.)$ is plotted for the three different electroweak bosons ($V = Z$, $W^+$, $W^-$ respectively), with a single set of flavour-independent (f.i.) non-perturbative parameters in the transverse part of the TMD PDFs and 50 different flavour-dependent (f.d.) sets of the same parameters. The values of the non-perturbative parameters have been chosen from the results in Ref.~\cite{Signori:2013mda,Bacchetta:2017gcc}.
}
\label{f:flav_ratios}
\end{figure}

We now analyse the ratios of the $q_\st$-differential cross section calculated with a flavour-independent set of non-perturbative parameters in $\widetilde{f}_{{\rm NP}}^a (b_\st; \bm{\lambda}^\prime)$ over the same cross section calculated with flavour-dependent parameters. The results are presented in Fig.~\ref{f:flav_ratios} for $Z$, $W^+$, $W^-$.  
The calculation has been performed by means of a flavour-dependent modification of {\tt DyqT}, where the non-perturbative contributions in Eq.~\eqref{e:TMDevol1} have been coded as:
\begin{equation}
\label{e:fd_dyqt}
\exp\{ -g_{NP}^a \} = \exp \{ -[ g_{evo} \ln(Q^2/Q_0^2) + g_a ] b_\st^2 \} \ .
\end{equation}
The values for $g_{evo}$, $Q_0$, $g_a$ are taken from Ref.~\cite{Bacchetta:2017gcc} and the flavour-dependence in $g_a$ is inspired to the flavour ratios in Ref.~\cite{Signori:2013mda}. The curves in Fig.~\ref{f:flav_ratios} correspond to 50 sets of flavour-dependent non-perturbative parameters built according to these criteria. The perturbative accuracy is NLL and the collinear PDF set used is NNPDF3.1~\cite{Ball:2017nwa}.

As predicted by the TMD formalism, the effect induced by the non-perturbative corrections is more evident at low $q_\st$. In particular, it is stronger for $q_\st < 5$ GeV but sizable up to $q_\st = 10$ GeV. The flavour dependence of the intrinsic transverse momentum can modify the shape of $d\sigma/dq_\st$ by $\sim 5-10 \%$ at very low transverse momentum. This observable affects the cross section differential with respect to the kinematics of the final state particles, namely the distributions in $p_\st^\ell$, $p_\st^\nu$, $m_\st$, and thus has an impact also on the determination of the $W$ boson mass.

\section{Impact on the determination of the $W$ boson mass}
\label{s:impact_wmass}

As previously mentioned, the measurements of $m_{W}$ at hadron colliders rely on a template-fit procedure performed on selected observables, i.e., the distributions in the transverse mass of the lepton pair and the lepton/neutrino transverse momentum. Both {\tt CDF} and {\tt D0} experiments at Tevatron use data from all the three observables. In the {\tt ATLAS} case, however, the transverse momentum of the (anti)neutrino is used for consistency checks only, since it is affected by larger uncertainties with respect to $m_{\st}$ and $p_\st^\ell$.

In this section we consider selected results concerning the estimate of the uncertainties of non-perturbative origin on the determination of $m_{W}$. In particular, we focus on shifts of the $W^\pm$ mass induced by possible configurations for the flavour dependence of the intrinsic transverse momentum, and we will compare them with the corresponding shifts generated by the uncertainties in the collinear PDFs.

In the template-fit procedure, several histograms are generated with a specific theoretical accuracy and description of detector effects, letting the fit parameter(s) (only $m_{W}$, in this case) vary in a range: the histogram best describing the experimental data selects the measured value for $m_{W}$. The details of the theoretical calculations used to compute the templates (the choice of the scales, of the collinear PDFs, of the perturbative order, the resummation of logarithmically enhanced contributions, the nonperturbative effects, etc.) affect the result of the fit and define the theoretical systematics.

\begin{figure}[hbt!]
\begin{center}
\includegraphics[width=0.5\textwidth]{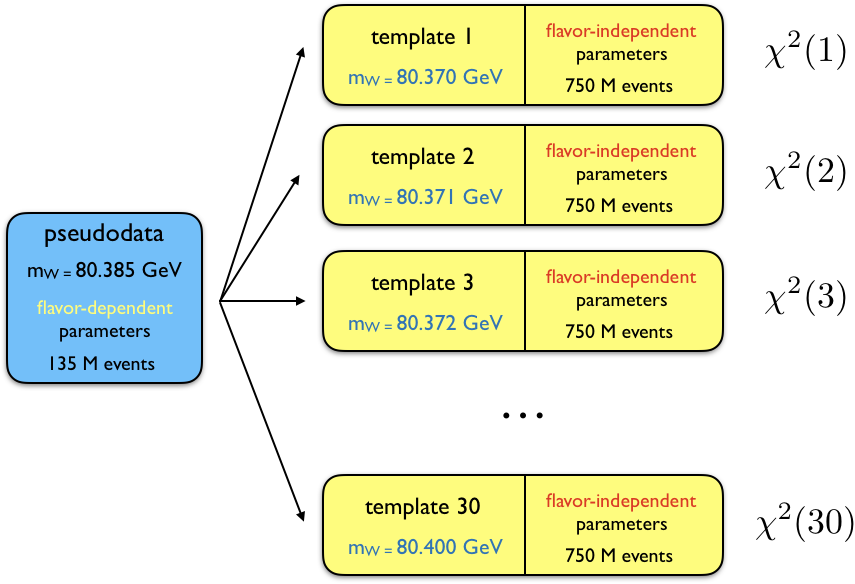} 
\caption{
Flowchart for a template-fit procedure to estimate shifts in $m_{W}$ induced by the flavour dependence of the intrinsic quark transverse momentum.
}
\label{f:template-fit}
\end{center}
\end{figure}

This procedure can also be used to estimate the effect of each {\it single} theoretical uncertainty, by generating sets of pseudodata (with the same event generator used for the templates, but at a lower statistics) differing by the value of the parameter(s) controlling that uncertainty~\cite{CarloniCalame:2003ux,CarloniCalame:2005vc}. Fig.~\ref{f:template-fit} contains a graphical illustration of the flowchart for the template-fit procedure, specified to the comparison of one set of pseudo-data generated with flavour-dependent parameters with 30 templates generated with one set of flavour-independent parameters and 30 values of $m_W$ ($80 385 \pm 15$ MeV with steps of 1 MeV). This method has been also used to estimate the shift in $m_{W}$ induced by the variation of the collinear PDF set in fitting the transverse mass~\cite{Bozzi:2011ww,Quackenbush:2015yra} and the lepton $p_\st$~\cite{Bozzi:2015hha,Quackenbush:2015yra} both at Tevatron and at the LHC in the central rapidity region of the produced electroweak boson ($|\eta|<1.0$ for Tevatron and $|\eta|<2.5$ for the LHC). A similar study dedicated to LHCb and its forward acceptance $2<\eta<4.5$ has been performed in Ref.~\cite{Bozzi:2015zja}.

\begin{figure}[hbt!]
\begin{center}
\includegraphics[width=0.4\textwidth]{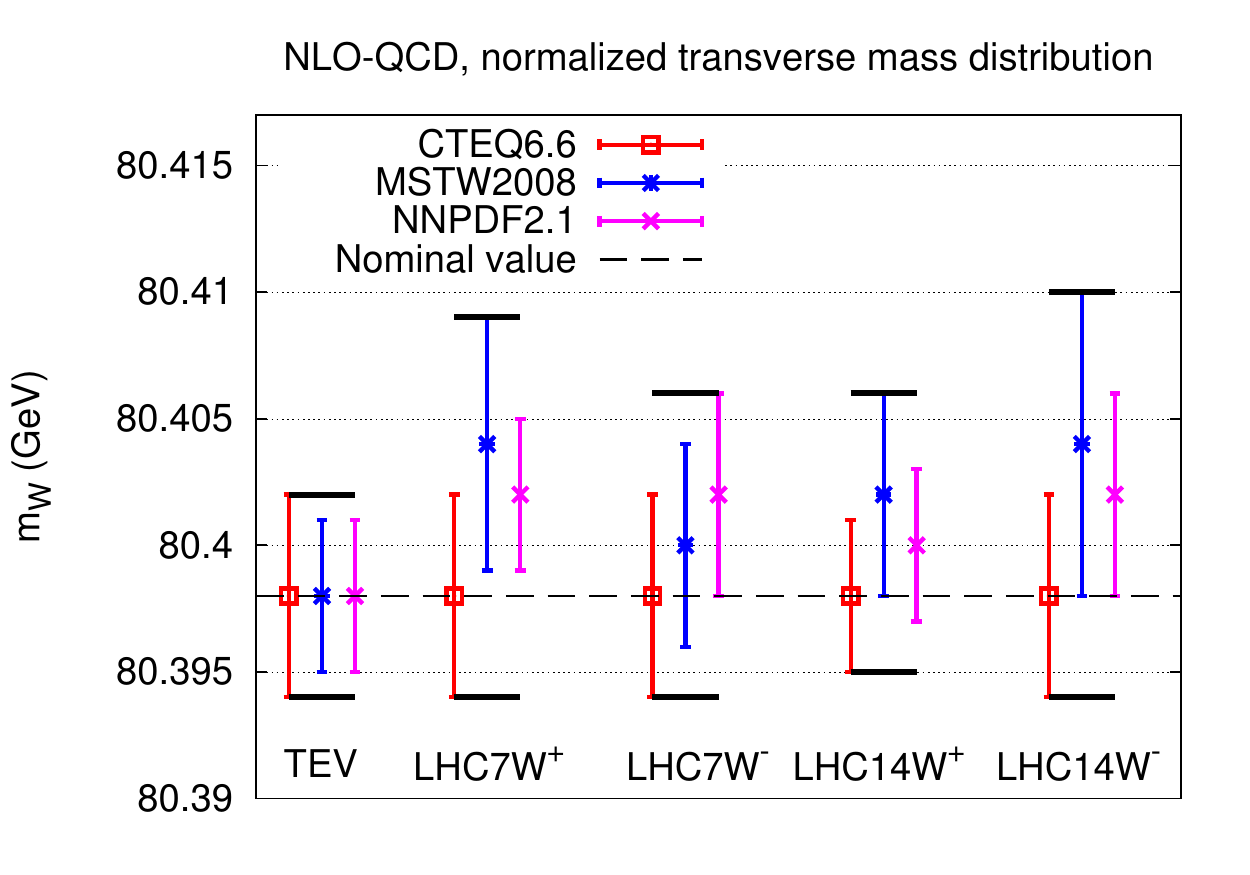}\;\;\;\;\;\;
\includegraphics[width=0.4\textwidth]{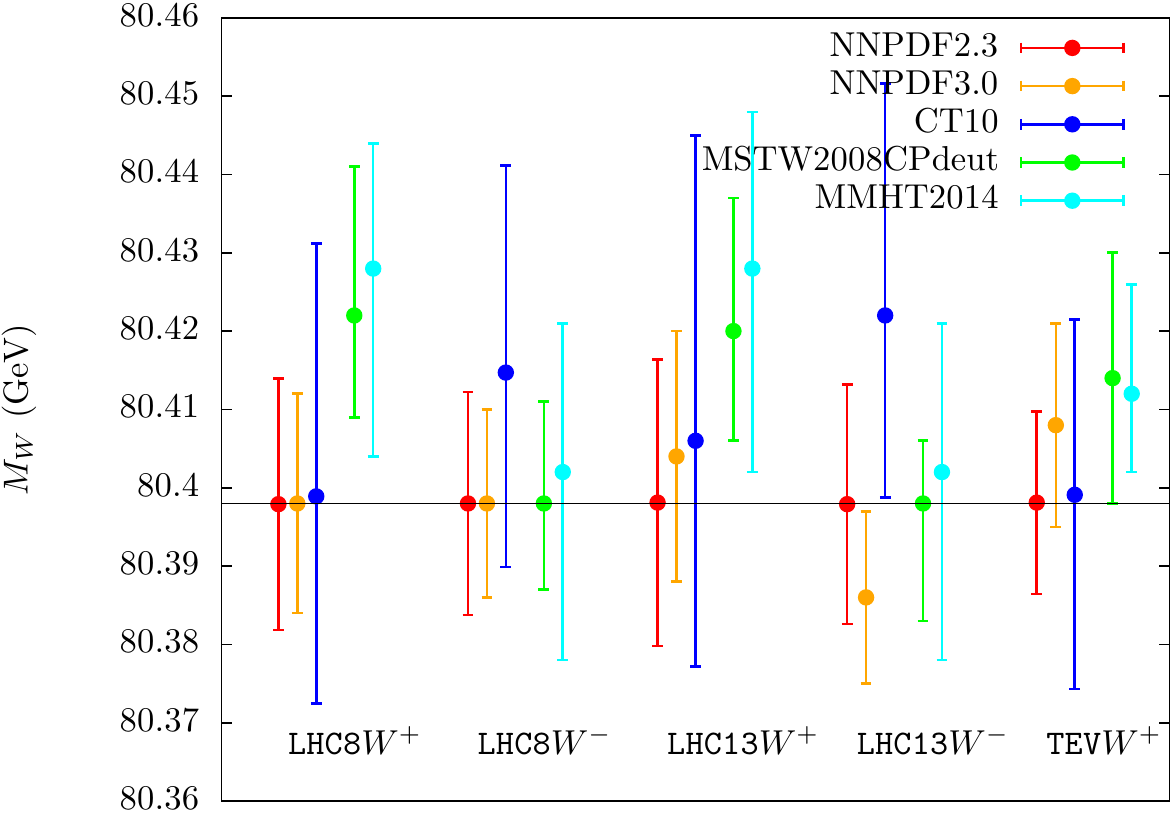} 
\caption{
Shifts induced on $m_{W}$ by the choice of different PDF sets, obtained through a template-fit performed on the transverse mass $m_\st$ (left) and the lepton $p_\st$ (right) observables (left figure from Ref.~\cite{Bozzi:2011ww}, right figure from Ref.~\cite{Bozzi:2015hha}).
}
\label{f:template-fit-results}
\end{center}
\end{figure}

In the transverse mass case, the total error (envelope) induced by three different PDF sets (CTEQ6.6~\cite{Nadolsky:2008zw}, MSTW2008~\cite{Martin:2009iq}, NNPDF2.1~\cite{Ball:2011mu}) is less than 10 MeV both at the Tevatron and at the LHC~\cite{Bozzi:2011ww}. The results are shown in the left plot of Fig.~\ref{f:template-fit-results}. The analysis has been performed at fixed-order NLO QCD (${\cal O}(\alpha_{s}$)), thus without all-order resummation, since the $m_\st$-shape is mildly sensitive to soft gluon emission from the initial state. The key factor in reducing the PDF uncertainty is the use of normalised differential distributions in the fitting procedure, in such a way to eliminate normalisation effects which are irrelevant for $m_{W}$. 

A similar analysis applied to the lepton $p_\st$ observable reveals a much larger error due to PDF variations (CT10~\cite{Gao:2013xoa}, MSTW2008CPdeut~\cite{Martin:2009iq}, MMHT2014~\cite{Harland-Lang:2014zoa}, NNPDF2.3~\cite{Ball:2012cx}, NNPDF3.0~\cite{Ball:2014uwa}), as shown in the right plot of Fig.~\ref{f:template-fit-results}. While the individual sets provide non-pessimistic estimates (${\cal O}(10\;\rm MeV)$), the distance between the best predictions of the various sets ranges between 8 and 15 MeV, and the total envelope ranges between 16 and 32 MeV (depending on the collider, the energy and the final state)~\cite{Bozzi:2011ww}. 
While soft gluon emission already provides a non-vanishing transverse momentum, additional contributions may come from the intrinsic transverse momentum of the colliding partons. The study of the impact of a possible flavour-dependent intrinsic $k_\st$ on the determination of $m_{W}$ has been first performed in Ref.~\cite{Bacchetta:2018lna}, using the same template-fit procedure described above and sketched in Fig.~\ref{f:template-fit}, performed with modified versions of the {\tt DYqT}~\cite{Bozzi:2010xn} and {\tt DYRes}~\cite{Catani:2015vma} codes. In this case, the pseudodata are built with the Gaussian widths $g_a$ associated to the different flavours in Eqs.~\eqref{e:fd_dyqt}. 

In order to estimate the impact of the flavour dependence, it is necessary to first identify the ``$Z$-equivalent'' sets of parameters, i.e., those sets in agreement with the $Z$ transverse momentum distribution measured at hadron colliders. To this extent:
\begin{itemize}
\item a single flavour-independent ({\em i.e.}, using a version of Eq.~\eqref{e:fd_dyqt} without $a$-dependence) $q_\st$-spectrum for the $Z$ boson  is produced based on the parameters presented in Ref.~\cite{Bacchetta:2017gcc};
\item each bin of this flavour-independent spectrum is assigned an uncertainty equal to the one quoted by the {\tt CDF} and {\tt ATLAS} experiments;
\item several flavour-dependent sets for $g_a$ in Eq.~\eqref{e:fd_dyqt} are generated randomly within a variation range consistent with the information obtained in previous TMD fits (in particular, taking into account the estimate for the flavour-independent contribution to the non-perturbative part of the evolution obtained in Ref.~\cite{Bacchetta:2017gcc});
\item a flavour-dependent set is defined ``$Z$-equivalent'' if the associated $q_\st$ spectrum for the $Z$ has a $\Delta\chi^{2}\leq 1$ with respect to one generated by the flavour-independent set.
\end{itemize}

The flavour-dependent sets for {\tt CDF} and {\tt ATLAS} who pass this filter are treated as the pseudodata of the template-fit procedure, while the flavour-independent one is used for the generation of the templates at high statistics. The number of events corresponds to 135M for the pseudodata and 750M for the templates. Only 9 sets out of the 30 ones which are ``$Z$-equivalent'' both with respect to {\tt CDF} and {\tt ATLAS} uncertainties have been investigated. The values of the flavour-dependent parameters for each set are given in Tab.~\ref{t:NPsets}. A summary of the shifts obtained through this procedure is given in Tab.~\ref{t:mw_shifts}.  

\begin{table}[htp]
\begin{tabular}{|c|c|c|c|c|c|}
\hline
Set & $u_{v}$ & $d_{v}$ & $u_{s}$ & $d_{s}$ & $s$ \\
\hline
1 & 0.34 & 0.26 & 0.46 & 0.59 & 0.32 \\
2 & 0.34 & 0.46 & 0.56 & 0.32 & 0.51 \\
3 & 0.55 & 0.34 & 0.33 & 0.55 & 0.30 \\
4 & 0.53 & 0.49 & 0.37 & 0.22 & 0.52 \\
5 & 0.42 & 0.38 & 0.29 & 0.57 & 0.27 \\
6 & 0.40 & 0.52 & 0.46 & 0.54 & 0.21 \\
7 & 0.22 & 0.21 & 0.40 & 0.46 & 0.49 \\
8 & 0.53 & 0.31 & 0.59 & 0.54 & 0.33 \\
9 & 0.46 & 0.46 & 0.58 & 0.40 & 0.28 \\
\hline
\end{tabular}
\caption{Values of the $g_{NP}^a$ parameter in Eq.~\ref{e:fd_dyqt} for the flavours $a=u_{v},d_{v},u_{s},d_{s},s=c=b=g$. Units are GeV$^2$.
} 
\label{t:NPsets}
\end{table}
\begin{table}[htp]
\small
\begin{tabular}{|c|cc|cc|}
  \hline
  \multicolumn{1}{|c|}{}&\multicolumn{2}{|c|}{$\Delta m_{W^+}$}&\multicolumn{2}{|c|}{$\Delta m_{W^-}$} \\
  \hline
  \hline
Set  & $m_\st$ & $p_\st^\ell$ & $m_\st$ & $p_\st^\ell$\\
\hline
1 & 0 & -1 & -2 & 3 \\
2 & 0 & -6 & -2 & 0 \\
3 & -1 & 9 & -2 & -4 \\
4 & 0 & 0 & -2 & -4 \\
5 & 0 & 4 & -1 & -3 \\
6 & 1 & 0 & -1 & 4 \\
7 & 2 & -1 & -1 & 0 \\
8 & 0 & 2 & 1 & 7 \\
9 & 0 & 4 & -1 & 0 \\
\hline
\end{tabular}
\caption{Shifts in $m_{W^\pm}$ (in MeV) induced by the corresponding sets of flavour-dependent intrinsic transverse momenta outlined in Tab.~\ref{t:NPsets} (Statistical uncertainty: 2.5 MeV).}
\label{t:mw_shifts}
\end{table} 

The statistical uncertainty of the template-fit procedure has been estimated by considering statistically equivalent those templates for which $\Delta \chi^2 = \chi^2 - \chi_{min}^2 \leq 1$. Overall, the quoted statistical uncertainty on the results in Tab.~\ref{t:mw_shifts} is $\pm 2.5$ MeV. 

Being the transverse mass mildly sensitive to the modeling of the $W^\pm$ transverse momentum, the corresponding shifts are compatible with zero considering the statistical uncertainty of the template-fit procedure. On the contrary, in the $p_\st^\ell$ case the shifts can be incompatible with statistical fluctuations and are comparable to the ones induced by collinear PDFs, with an envelope of 15 MeV in the case of $W^{+}$ production and 11 MeV for $W^-$ production. We also notice a hint of a possible anti-correlation between the shifts in the $W^{+}$ and $W^{-}$ cases, as  it was also noticed in Sec.~\ref{s:effects_peak_wmass}. 

Along this line, we also stress that {\tt ATLAS} measured $m_{W^+} - m_{W^-} = -29 \pm 28$ MeV~\cite{Aaboud:2017svj}. From Tab.~\ref{t:mw_shifts}, we can infer that part of the discrepancy between the mass of the $W^+$ and the $W^-$ can be artificially induced by not considering the flavour structure in transverse momentum. For example, the sets 1 and 2 in Tab.~\ref{t:NPsets} feature $\delta m_{W^-} > \delta m_{W^+}$ (induced by $p_\st^\ell$). This implies that for templates built with sets 1 and 2, instead of flavour-independent values, the difference between the two masses would be reduced. An opposite result would be obtained if building templates with flavour-dependent sets for which $\delta m_{W^-} < \delta m_{W^+}$ (e.g. sets 3 and 5, for the $p_\st^\ell$ case). 

\section{Outlook and future developments}
\label{s:future_dev_wmass}

The selected results presented in this contribution point out that the impact of a possible flavour dependence of the intrinsic partonic transverse momentum should not be neglected, even in the kinematic region where nonperturbative effects are expected to be small~\cite{Berger:2002ut,Berger:2003pd,Berger:2004cc}, such as for electroweak boson production at the LHC.

This kind of uncertainty directly affects the electroweak observables relevant for the measurement of $m_{W}$: the transverse momentum distribution for the $W$ and the decay lepton, and the transverse mass  distribution of the lepton pair. The numerical results presented in Sec.~\ref{s:effects_peak_wmass} and Sec.~\ref{s:impact_wmass} indicate that flavour-dependent effects are comparable in size to other uncertainties of (non-)perturbative origin (for example, the choice of collinear PDF set). Thus, a flavour-blind analysis is not a sufficiently accurate option for a program of precision electroweak measurements at the LHC and at future colliders.

Moreover, in hadron colliders at a lower energy such as RHIC and a possible fixed-target experiment at the LHC, the non-perturbative effects can play an even more significant role (due to the larger $x$-values probed) and affect the study of polarised TMDs~\cite{Aschenauer:2015ndk} and the structure of the light sea quarks~\cite{Hadjidakis:2018ifr}.

A detailed knowledge of TMD distributions is thus important, not only for nucleon tomography beyond the collinear picture~\cite{Boer:2011fh,Dudek:2012vr,Accardi:2012qut,Angeles-Martinez:2015sea,Diehl:2015uka,Boglione:2015zyc,Bacchetta:2016ccz,Boer:2016xqr,Metz:2016swz}, but also to constrain fundamental parameters of the Standard Model, thus providing a direct connection between hadron physics and the high-energy phenomenology.

In light of these results, we call for improved investigations of the impact of nonperturbative effects linked to the hadron structure at hadron colliders and for the inclusion of these effects in the event generators employed in experimental and theoretical investigations of high-energy physics.


\begin{appendix}
\section{Conventions for nonperturbative parameters}
\label{app:con_NP_par}

For convenience, we collect in this Appendix the naive translation of the nonperturbative parameters used in the numerical codes cited in the text. In the conventions of~\cite{Signori:2013mda,Bacchetta:2015ora}, the nonperturbative parameters appear as:
\begin{equation}
	\dd \sigma \propto \exp \left( - \frac{1}{4} \left( \langle k_T^2\rangle_{q_1} + \langle k_T^2 \rangle_{q_2} \right) b_T^2 \right) \ .
\end{equation}
In \href{http://cute.hepforge.org/}{CuTe}~\cite{Becher:2011xn} there is a single nonperturbative parameter entering the cross section:
\begin{equation}
	\dd \sigma \propto \exp \left( - 2 \Lambda_{NP}^2 b_T^2 \right) \ .
\end{equation}
The same happens in \href{http://pcteserver.mi.infn.it/~ferrera/dyqt.html}{DyqT}~\cite{Bozzi:2010xn} and \href{http://theory.fi.infn.it/grazzini/dy.html}{DYRes}~\cite{Catani:2015vma}, in terms of the nonperturbative parameter $g_{NP}$:
\begin{equation}
\dd \sigma \propto \exp \left( - g_{NP} b_T^2 \right) \ .
\end{equation}
We obtain the parameter employed in CuTe as
\begin{equation}
\begin{split}
\Lambda_{NP} &= \sqrt{ \frac{1}{8} \left( \langle k_T^2\rangle_{q_1} + \langle k_T^2 \rangle_{q_2} \right)} \ ,\\
\Lambda_{NP} &= \sqrt{ g_{NP}/2 } \ .
\end{split}
\end{equation}
and similarly for the DYqT parameter:
\begin{equation}
\label{e:rel_par}
\begin{split}
	g_{NP} &= \frac{1}{4} \left( \langle k_T^2\rangle_{q_1} + \langle k_T^2 \rangle_{q_2} \right) \ , \\
	g_{NP} &= 2 \Lambda_{NP}^2 \ .
\end{split}
\end{equation}
Here we report the most important values, based on Eq.~\eqref{e:rel_par}:
\begin{table}[h!]
\begin{tabular}{llll}
&	                              $\Lambda_{NP}$ 	            &	$g_{NP}$   	      & $\langle k_T^2 \rangle (q_1 = q_2)$ \\
CuTe default				      & 0.60					    & 0.72                & 1.44     \\
DYqT conservative estimate\ \ \	  & 0.77\ \ \ \ \ \				& 1.2\ \ \ \ \ \      & 2.40
\end{tabular}
\end{table}

\end{appendix}


\begin{acknowledgments}
We thank Alessandro Bacchetta and Marco Radici for the fruitful collaboration on this topic, Alessandro Vicini for many suggestions and discussions, Chao Shi for carefully reading the manuscript and Piet Mulders and Mathias Ritzmann for contributing to the investigations summarized in this article. AS acknowledges support from the U.S. Department of Energy, Office of Science, Office of Nuclear Physics, contract no. DE-AC02-06CH11357. GB acknowledges support from the European Research Council (ERC) under the European Union's Horizon 2020 research and innovation program (grant agreement No. 647981, 3DSPIN).
\end{acknowledgments}


\bibliographystyle{JHEP}  
\bibliography{biblio_wunc}

\end{document}